\documentclass[aps,prb,twocolumn]{revtex4-1}
\usepackage{amssymb}
\usepackage{graphicx}
\usepackage{hyperref}

\begin{document}
\title{ Density wave like transport anomalies in surface doped Na$_{2}$IrO$_3$}.
\author{Kavita Mehlawat and Yogesh Singh}
\affiliation{Indian Institute of Science Education and Research (IISER) Mohali, Knowledge city, Sector 81, Mohali 140306, India}

\date{\today}

\begin{abstract}
We report that the surface conductivity of Na$_{2}$IrO$_3$ crystal is extremely tunable by high energy Ar plasma etching and can be tuned from insulating to metallic with increasing etching time. Temperature dependent electrical transport for the metallic samples show signatures of first order phase transitions which are consistent with charge or spin density wave like phase transitions recently predicted theoretically. Additionally, grazing-incidence small-angle x-ray scattering (GISAXS) reveal that the room temperature surface structure of Na$_{2}$IrO$_3$ does not change after plasma etching.

\end{abstract}

\maketitle

\section{Introduction}
\label{sec:INTRO}

5d transition metal oxides (TMO) are candidates for novel magnetism \cite{Jackeli2009, Crawford1994}. In 3d TMOs, electron correlation (U) is the most significant energy scale. However, for 5d TMOs such as Ir-based oxides, spin-orbit coupling (SOC) becomes comparable to U and plays an important role in determining the electronic band structure and magnetism.  For example, the novel $J_{eff} $ = 1/2 state in the perovskite iridate Sr$_{2}$IrO$ _{4} $ arises from splitting of the t$ _{2g}$ level due to spin-orbit coupling \cite{Kim2009, Kim2008}. Sr$_{2}$IrO$ _{4}$ has garnered a lot of recent attention due to its structural, electronic and magnetic similarities with La$_{2}$CuO$ _{4} $  \cite{Crawford1994, Kim2008}. Recently $in situ$ surface electron doping by potassium deposition has been achieved, and it was found that Fermi arcs, pseudogap, and a low-temperature d-wave gap exist for these samples demonstrating properties in complete analogy with the cuprates \cite{Kim2014, Kim2016, Yan2015}. These studies on Sr$_2$IrO$_4$ suggest that properties of other iridates might also be highly tunable. Recently the layered honeycomb lattice Iridates A$_{2}$IrO$ _{3} $ (A = Na, Li) have been recognized as spin-orbit assisted $J_{eff} $ = 1/2 Mott insulators \cite{ Jackeli2009, Singh2010, Choi2012, Singh2012}. The magnetic frustration observed in these materials is argued to arise from the presence of dominant bond-directional Kitaev-like exchange interactions with additional Heisenberg exchange present \cite{Chun2015,Singh2012}.  Recent predictions have been made of novel states emerging on doping the Kitaev-Heisenberg model. These include topological superconductivity \cite{You2012, Okamoto2013}, spin/charge density waves, electronic dimerization instabilities, and bond order instabilities \cite{Scherer2014}. This motivated us to explore the properties of doped Na$_2$IrO$_3$ crystals.  

Conventional doping in the bulk has not been successful so far. We have however, recently discovered that the surface of these crystals can be doped by Argon plasma etching \cite{Mehlawat2016}. In this paper, we reproduce surface doping of Na$_2$IrO$_3$ crystals using reactive Ar ion etching and report temperature and magnetic field dependent electrical transport studies on these RIE treated Na$_2$IrO$_3$ crystals.  We find anomalies in the electrical transport properties which are consistent with spin/charge density wave-like phase transitions or structural transitions in these surface doped Na$_{2}$IrO$_{3}$ crystals.  

\section{EXPERIMENTAL DETAILS}
\label{sec:EXPT}
The single crystals of Na$_{2}$IrO$ _{3} $ were grown using self-flux growth method described in detail elsewhere \cite{Singh2010}. Reactive-ion etching (RIE) technique was used to modified the surface of the crystals. The freshly cleaved (in air) crystals were bombarded with high energy Ar plasma (in vacuum) for varying periods of time ranging from $0$ to $30$~min. The parameters used during plasma etching are given in Table~\ref{Table-Etching-parameters}. The surface structure of the crystals before and after RIE was checked using grazing incidence small angle x-ray scattering (GISAXS). The transport measurement have been done using a Quantum Design physical property measurement system.  The etched surface is found to be very sensitive to ambient lab environment. The change in the transport behaviour of etched samples is not permanent and reverts back to insulating behaviour on exposure to ambient atmosphere. The lab exposure time required for the samples to revert back to their original behaviour depends on the time they were etched for. For example, samples etched for longer time $\backsim 30$~min and $40$~min, degrade very fast, within about two hours or so.  Whereas, the crystals etched for $10$~min degrade in 1 to 2 days. Additionally, it was found that transport properties of the crystal surface opposite to the etched surface remains unchanged.  These observations suggest that only the surface of the Na$_{2}$IrO$_{3}$ crystal is modified and most likely only a small depth close to the top of the etched surface is affected.  Unfortunately the depth of the surface layer affected by the RIE is unknown.  For this reason we present electrical transport data as sheet resistance. 
  
\begin{table}
\caption{Summary of Parameters  used in high energy argon plasma etching}. 
\begin{ruledtabular}
\begin{tabular}{|cc|}
flow of Ar gas & 80 cubic centimeter per minute at\\ %\hline  
&standard temperature and pressure \\ %\hline  
Chamber pressure  &  80 mTorr  \\ %\hline
RF power & 200 W   \\ %\hline
RF bias voltage & -500 V   \\ %\hline 
Temperature & 15 $^\circ$C - 20 $^\circ$C   \\ %\hline 
\end{tabular}
\end{ruledtabular}
\label{Table-Etching-parameters}
\end{table} 
 
\section{RESULTS}
\subsection{Crystal Structure and Chemical Analysis}
\begin{figure}[t]   
\includegraphics[width= 3 in]{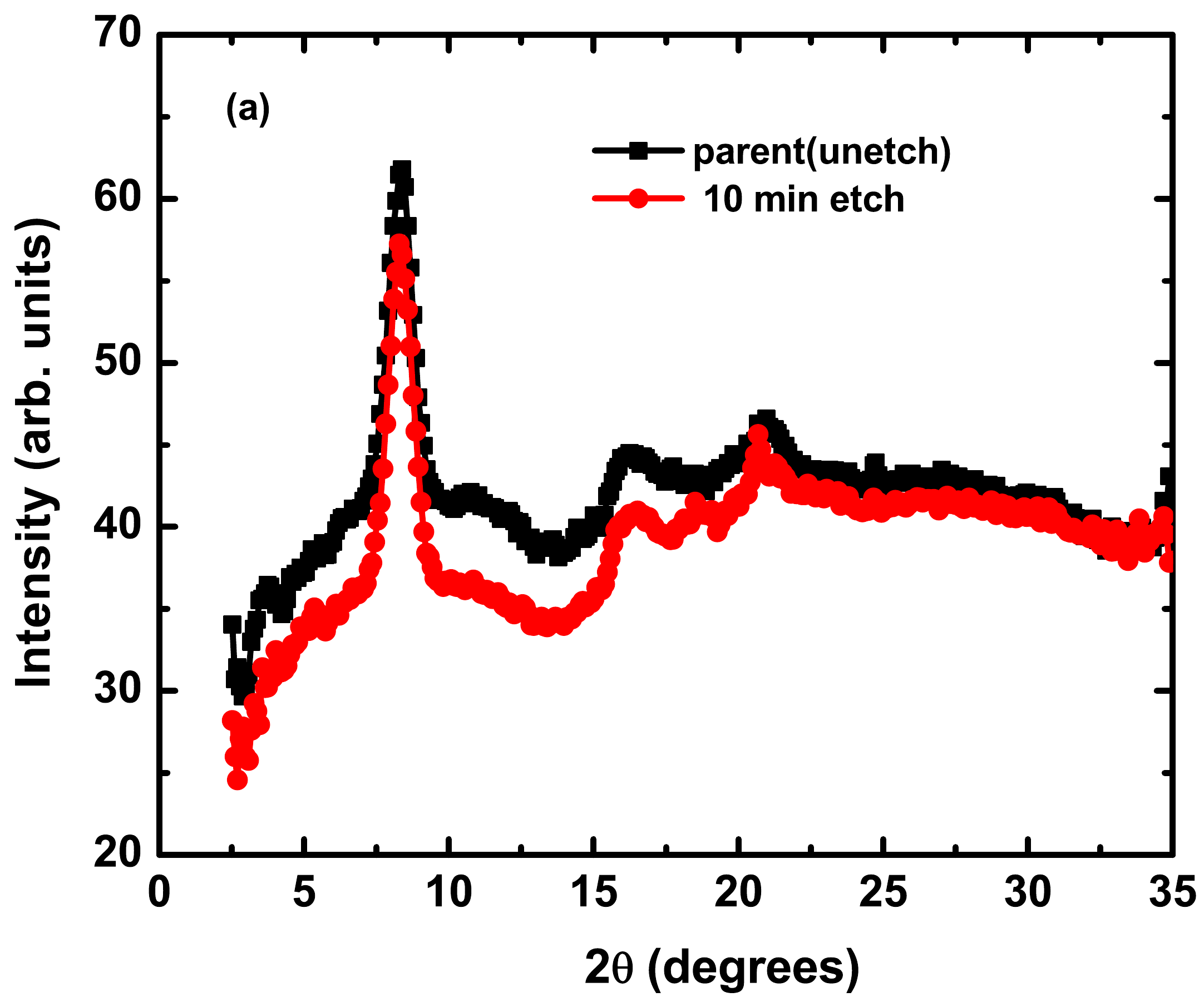}  
\includegraphics[width= 3 in]{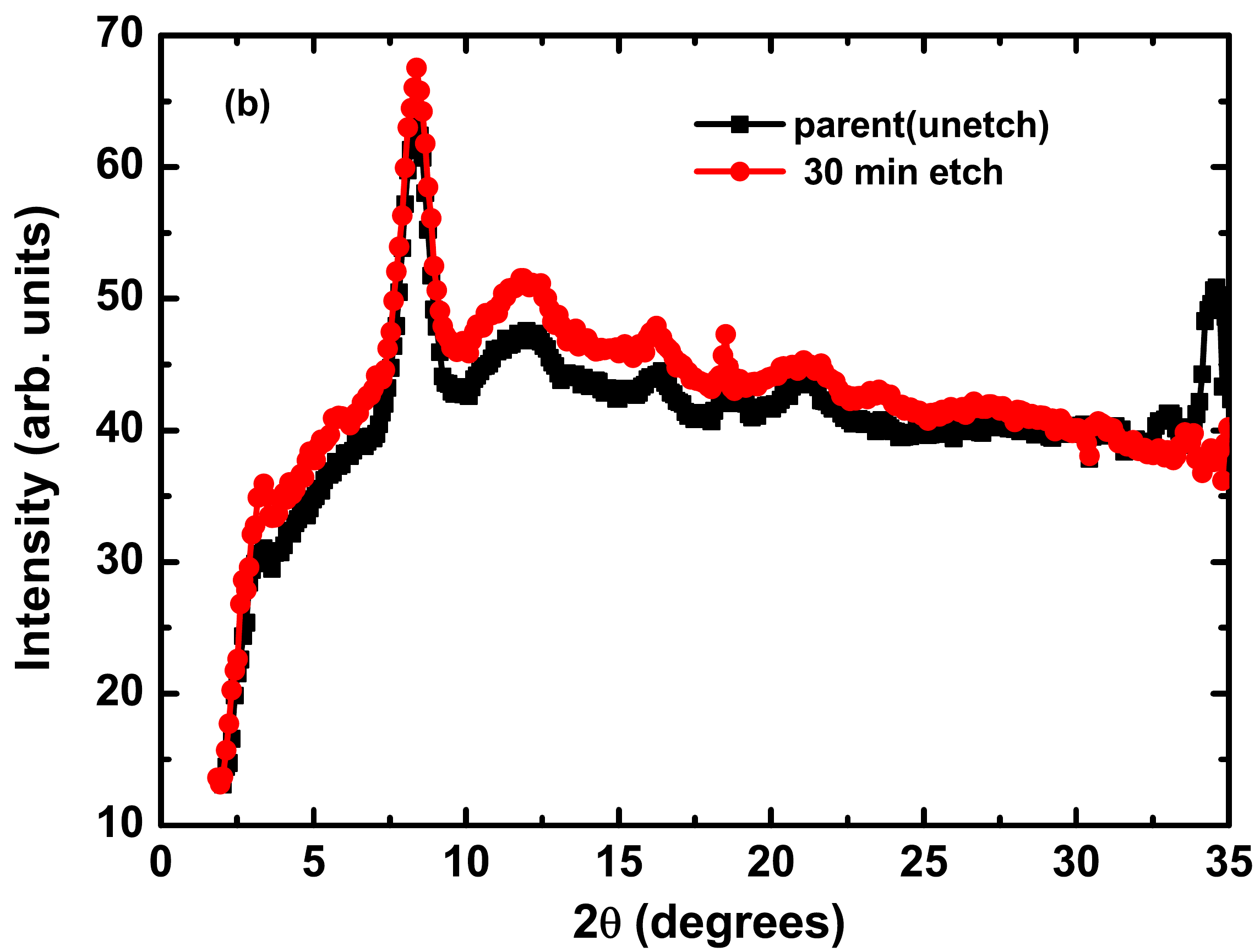}   
\caption{(Color online) The grazing incidence small-angle x-ray scattering patterns for Na$_{2}$IrO$ _{3} $ before and after  Ar plasma etching for 10 minutes and 30 minutes plasma etched crystals. 
\label{Fig-1}}
\end{figure}

The  grazing incidence small angle x-ray scattering (GISAXS) were done on the crystals before and after the plasma etching. Results of these measurements are displayed in Fig.~\ref{Fig-1}.  GISAXS reveals that the overall surface structure of the crystals does not change after plasma etching.  A notable change is for the high angle peak (32$^\circ$ - 34$^\circ$) which shifts to a smaller angle in the plasma etched samples.  This suggests an increase in the unit cell parameters.  This however, could not be quantified with the current data. The depth $D$ probed during the GISAXS measurement can be estimated using the relation $D = d$sin$\theta$, where $d \thickapprox $ 1.25$ \times $10$ ^{-3}$~cm is the attenuation length for this sample at the used X-ray energy of $15$~keV, and $ \theta $ = 0.1$^\circ$ is the x-ray incident angle measured from the surface of the crystal.  The estimated depth comes out to be $D \thickapprox 20$~nm confirming that GISAXS is probing the surface structure of the samples. Since the thickness of the surface layer modified by the RIE is unknown and probably depends on the exposure time of plasma etching, $D$ estimated above can be used as a lower limit.  The chemical composition of the etched surface was determined using the energy dispersive X-ray spectroscopy (EDS) on a JEOL scanning electron microscope (SEM). The results from this analysis are shown in Table~\ref{Table-EDS-parameter}. From these results, we can conclude that Na is being progressively removed from the surface of the samples with increasing exposure time to the plasma, which leads to hole doping at surface of the crystal. 

\begin{table}
\caption{Average chemical composition of the expected elements from energy dispersive x-ray spectroscopy}. 
\begin{ruledtabular}
\begin{tabular}{|ccc|}
Exposure time (min) & Average (Na) & Average (Ir) \\ \hline  
0  &  1.78 & 1 \\ %\hline
10 & 1.73  & 1 \\ %\hline
20 & 1.68 & 1  \\ %\hline 
30 & 1.62 & 1  \\ %\hline 
\end{tabular}
\end{ruledtabular}
\label{Table-EDS-parameter}
\end{table}  

\begin{figure}[t]   
\includegraphics[width= 3 in]{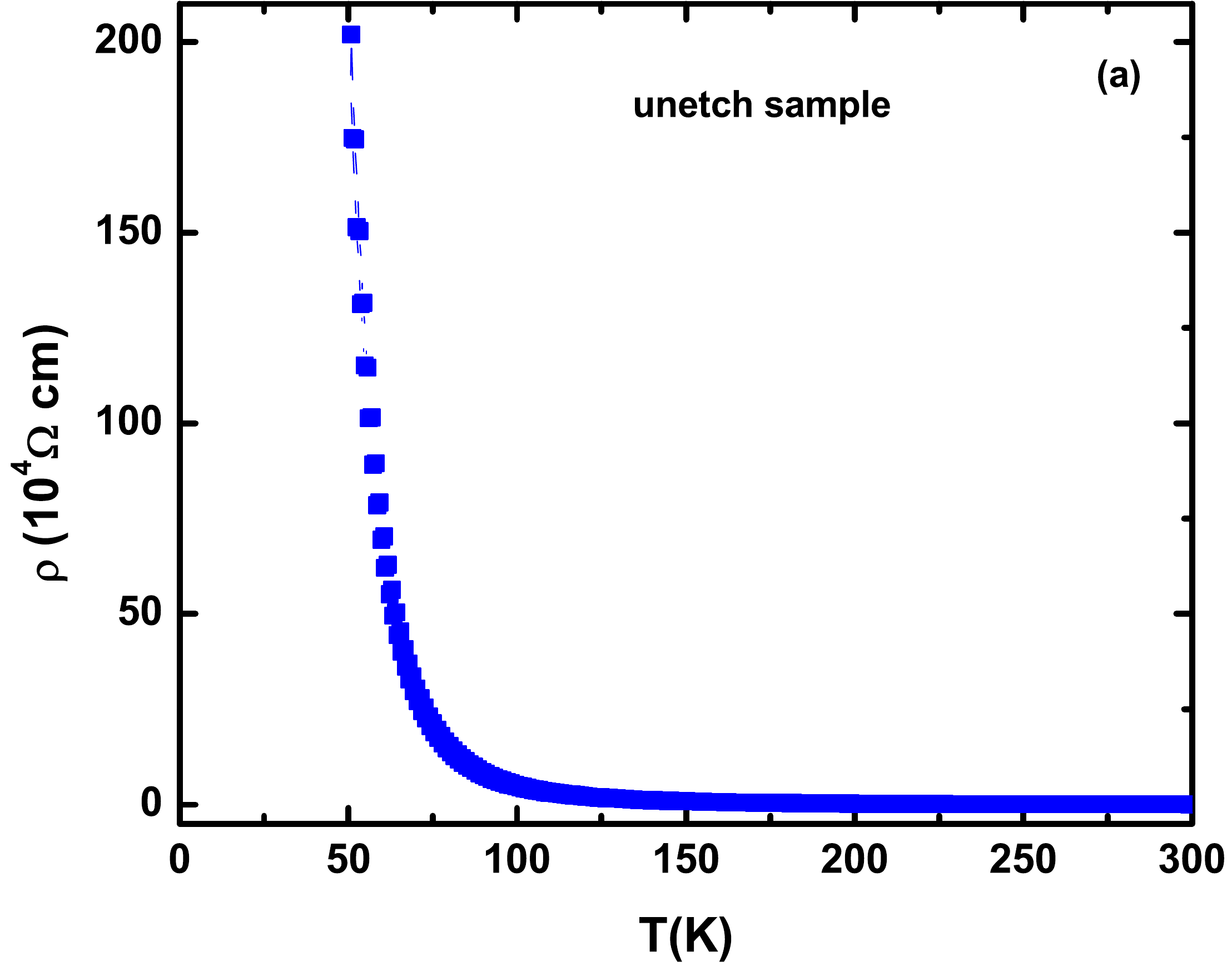}
\includegraphics[width= 3 in]{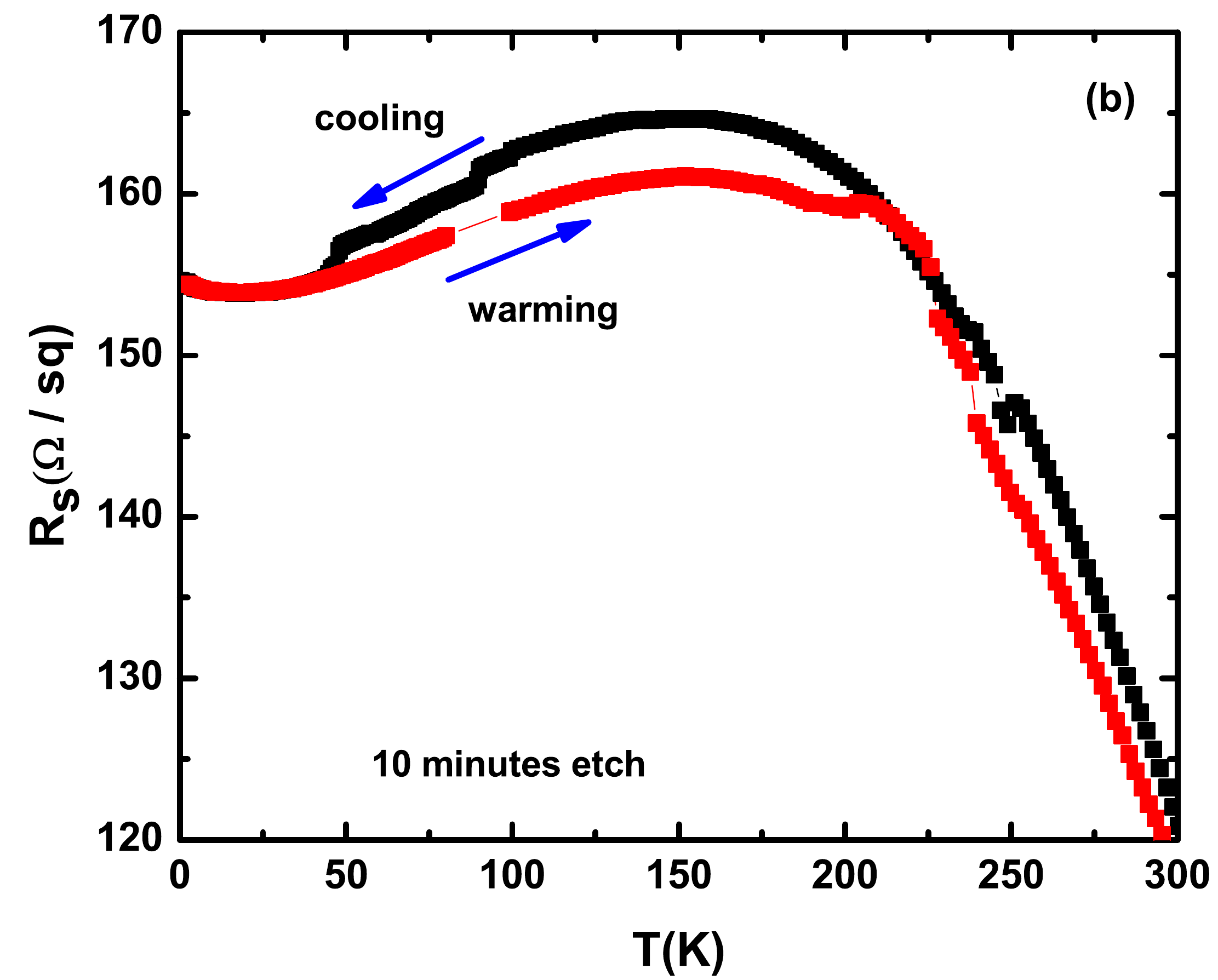}      
\caption{(Color online)The electrical resistivity $ \rho $ versus temperature T for unetch and electrical transport as sheet resistance R$ _{s} $ 10 minutes of plasma etched of the  Na$_{2}$IrO$ _{3} $ crystals at zero magnetic field.
\label{Fig-2}}
\end{figure}

\begin{figure}[h]   
\includegraphics[width= 3 in]{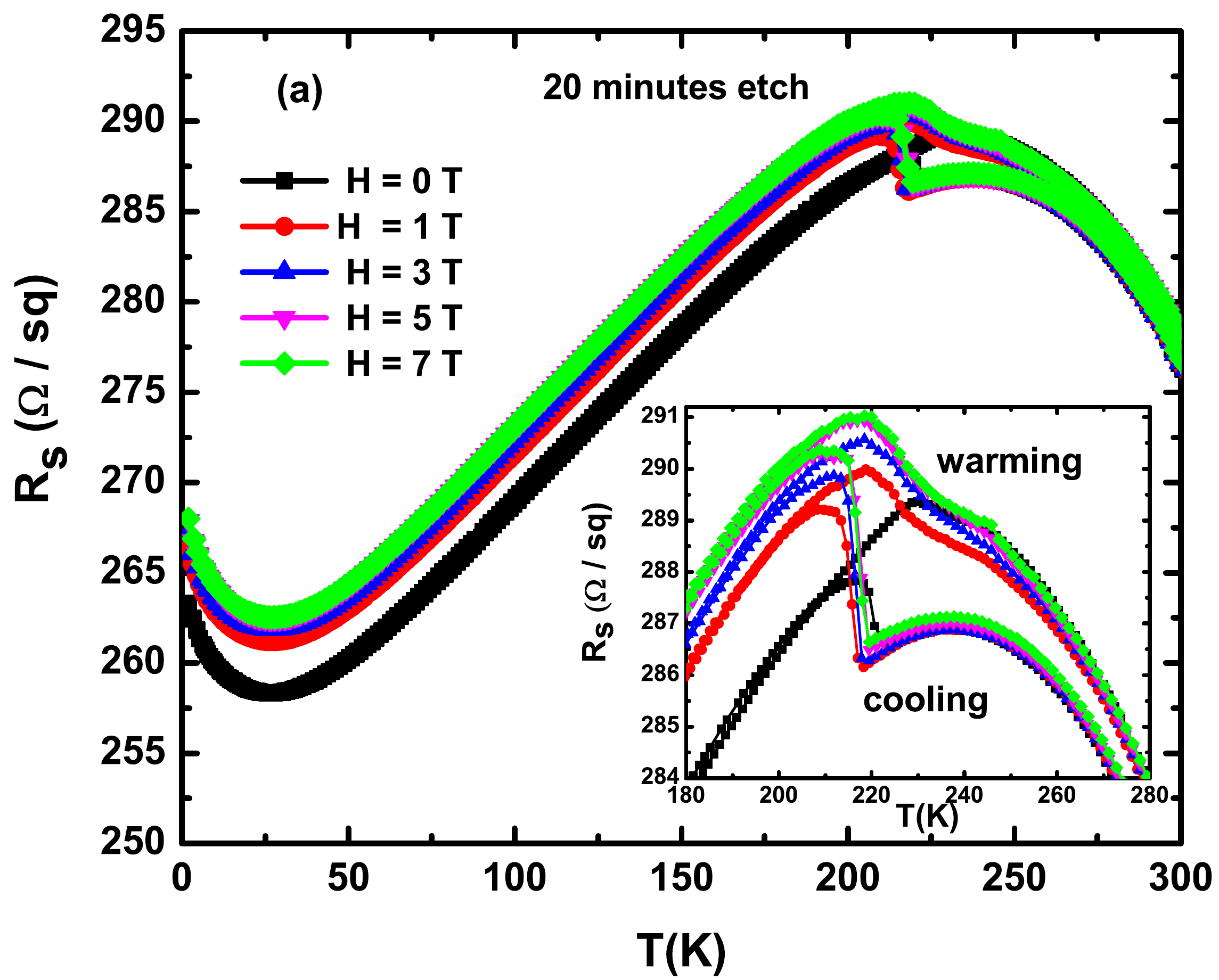}
\includegraphics[width= 3 in]{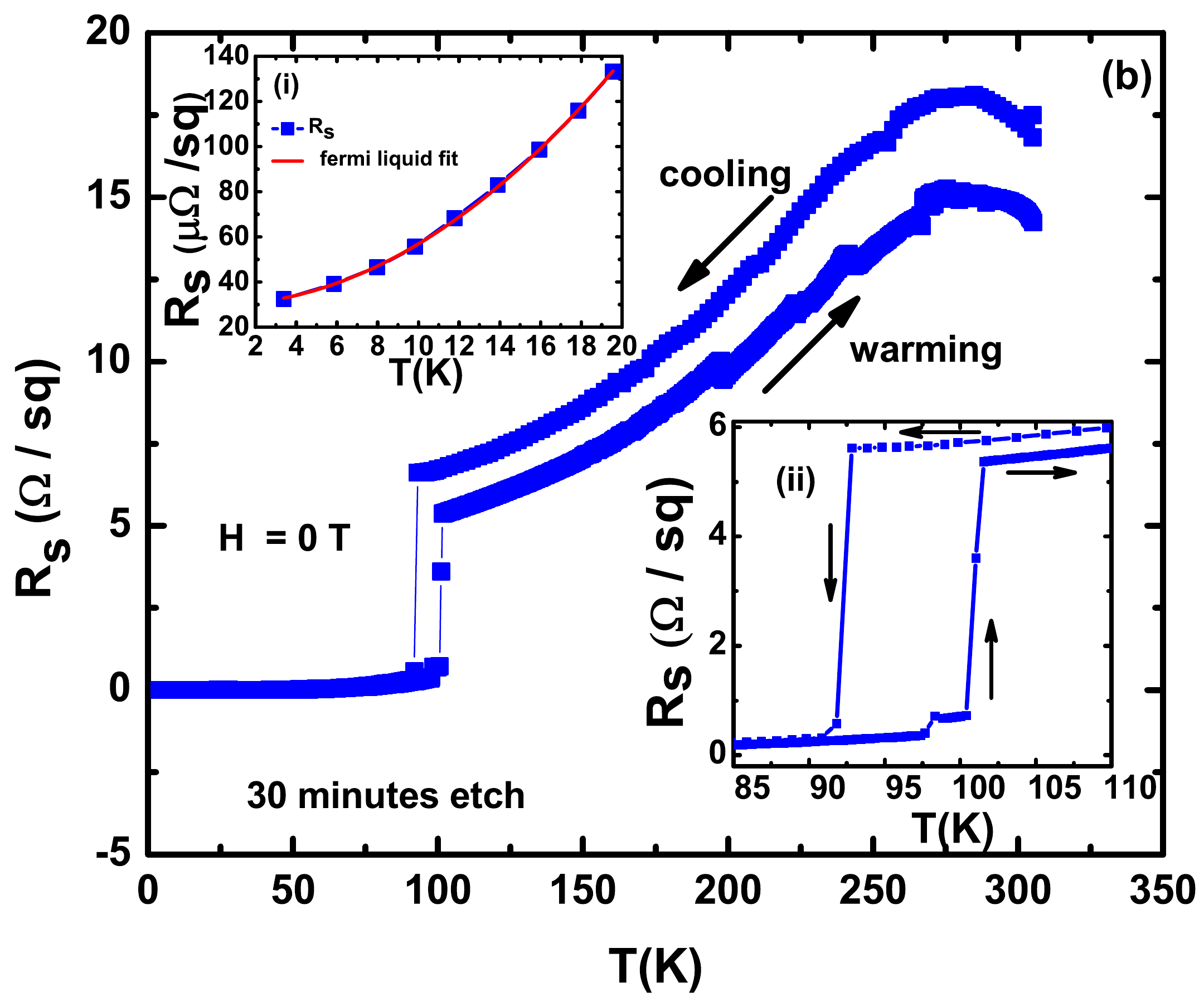}      
\caption{(Color online) The electrical transport as sheet resistance R$ _{s} $ versus temperature T for 20 minutes and 30 minutes of plasma etched Na$_{2}$IrO$ _{3} $ crystals. (a) R$ _{s} $ versus T measured in different
applied magnetic field H while cooling from T = 305 K and warming from T = 2 K. The inset shows
the cooling and warming data of different applied magnetic field H to highlight the thermal hysteresis which indicating the first-order nature of the phase transition. (b) R$ _{s} $ versus T  for 30 min exposure time crystal measured in a zero magnetic field H while cooling from T = 305 K and warming from T = 2 K. Inset(i) shows the low-temperature data fit to  a T$ ^{2} $ below T = 20 K suggesting Fermi-Liquid behavior. Inset(ii) shows the cooling and warming data at zero magnetic field H to highlight the thermal hysteresis indicating the first-order nature of the phase transition.  
\label{Fig-3}}
\end{figure}

\subsection{Electrical Transport}
Due to the unknown geometrical factors involved in the measurements, we present electrical transport for the etched samples as sheet resistance $R_s$ in the units $ \Omega $ / sq.  A square sheet with a sheet resistance of 10~$\Omega$/sq will have a resistance $ 10~\Omega $. 

The sheet resistance $R_s$ versus temperature $T$ data of the pristine (unetched) samples and the samples etched for various time periods, are shown in the Fig.~\ref{Fig-2} and Fig.~\ref{Fig-3}. Figure~\ref{Fig-2}~(a) show the $\rho$ versus $T$ data of the unetched Na$_{2}$IrO$ _{3}$ single crystal between $T = 50$ and $305$~K\@. The $\rho(T)$ data confirms the insulating behavior in the Na$_{2}$IrO$ _{3}$ sample as has been reported earlier \cite{Singh2010}. Figure~\ref{Fig-2}~(b) shows the sheet resistance $R_{s}$ versus $T$ data for the $10$~minutes etched crystal recorded while cooling the sample down from $T = 305$~K to $2$~K and while warming back up again. The $R_{s}$(T) versus $T$ increases on cooling from $T = 305$~K and reaches a broad maximum around $T = 100$-$200$~K\@. Below the maximum $R_{s}$(T) behaves like a metal with a weak $T$ dependence. This unusual behavior of transport is consistent with expectation for topological insulators where the insulating bulk dominates the transport at high temperature and the surface states start contributing at lower temperatures when the bulk gets gaped out. Such a topological state has been predicted for the Na$_{2}$IrO$_{3}$ under certain conditions \cite{Shitade2009, Kim2013}. The slight hysteresis in the warming and cooling curve is not understood at present.  We note that at the lowest temperatures there is a small increase in the resistance.  We believe that this weak upturn is arising from weak localization of the doped carriers due to the disorder introduced by the RIE process.  A non-monotonic electrical transport behaviour is also sometimes observed due to a crossover from a larger intrinsic gap at high temperatures to a smaller dopant gap at low temperatures in doped semiconductors or insulators.  A prominent example of this is the correlated insulator NiS$_2$ \cite{Yao1996}.  Figure~\ref{Fig-3}~(a) shows R$_{s}$(T) data measured for the sample etched for $20$~minutes, while cooling and warming in various applied magnetic fields $H$. The sheet resistance increases on cooling below T = 305 K and reaches a maximum around 250 K. A step-like increase iis observed n the $R_{s}$(T) at $T _{0} = 220$~K after which there is a rapid decrease in sheet resistance suggesting metallic behaviour down to 20 K, below which a slight increase in sheet resistance is observed again.  This upturn is stronger than observed for the $10$~min sample and will be consistent with a stronger localization due to the increased disorder introduced due to etching for longer times.  

The R$_{s}$(T) behavior close to the transition is highlighted in the inset of Fig.~\ref{Fig-3}~(a).  At T$ _{0}$ the step-like increase in R$_{s}$(T) is similar to signatures observed in electrical transport across charge-density-wave (CDW) phase transitions. In the case of a CDW the increase in R$_{s}$(T) indicates a partial loss of the density of states due to partial gaping of the Fermi surface \cite{Gruener2000}.  The transition temperature at T$ _{0} $ also shows thermal hysteresis of $ \thickapprox 10$~K between cooling and warming data and indicates a first order phase transition. The transition at $T_0$ is robust in applied magnetic field of up to $H = 9$~T as can be seen in the Fig.~\ref{Fig-3}~(a).  The magnitude of the increase in R$_{s}$(T) increases in applied magnetic fields as seen in the inset of ig.~\ref{Fig-3}~(a). Figure~\ref{Fig-3}~(b) show the R$_{s}$(T) versus T data for the sample etched for 30 minutes, measured in zero magnetic field while cooling and heating.  The sheet resistance increases on cooling below T = 305 K and reaches a maximum around 250 - 270 K, after which it continuously decreases down to T$ _{0} = 95$~K\@.  At T$ _{0} $ an abrupt step-like decrease in R$_{s}$(T) by more than an order of magnitude is observed and there is again a thermal hysteresis between cooling and warming data indicating the first order nature of the transition. The R$_{s}$ data at low temperatures follows a $T^2$ dependence indicating Fermi liquid behaviour.  The $T^2$ fit through the data is shown in the inset in Fig.~\ref{Fig-3}~(b).  If the complete crystal thickness ($\approx 0.1$~mm) is used in calculating the geometrical factor for converting resistance into resistivity $\rho$ then the coefficient of the $T^2$ dependence is estimated to be $A = 27~\mu \Omega$~cm/K$^2$.  This value is comparable to values obtained for Ce-based heavy Fermion compounds \cite{Stewart-RMP}.  

\section{SUMMARY AND DISCUSSION}
We have successfully shown that Ar plasma etching can be used to effectively dope the surface of Na$_{2}$IrO$_{3}$ crystals and that their electrical conductivity can be varied all the way from insulating to metallic by controlling the etching time. The surface structure of Na$_{2}$IrO$ _{3} $ does not change by plasma etching as revealed by grazing incident small angle x-ray scattering (GISAXS). The metallic samples are observed to show transport signatures consistent with those observed for density wave or structural transitions. Specifically, temperature dependent sheet resistance R$ _{s} $(T) for 20 and 30 minutes etched samples show charge density wave like phase transitions with an abrupt change in R$ _{s} $ at T$ _{0} $ = 220 and 95 K, respectively as shown in the Fig.~\ref{Fig-3}.  Such transport anomalies have been observed previously for CDW transitions in many other materials \cite{Yang1991, Singh2005, Singh2005a}.  The phase transition is first order as revealed by a 10 K thermal hysteresis between cooling and warming measurements. The most metallic (30 minutes etched) sample follows a T$^{2}$ Fermi liquid behavior at low temperature. Remarkably such spin/charge density waves, electronic dimerization instabilities, and bond order instabilities have been predicted on doping for antiferromagnetic Kitaev and ferromagnetic Heisenberg interaction\cite{You2012, Okamoto2013, Scherer2014}.  However, surface sensitive probes like electron diffraction or scanning tunnelling microscopy (STM) will be required to confirm the coupled electron density and structural modulations expected below a CDW transition \cite{Gruener2000}.  

\paragraph{Acknowledgments.--} We thank the small angle x-ray scattering facility at IISER Mohali. We thank Dr. A. Venkatesan for use of the plasma etching facility and we acknowledge the SEM facility at IISER Mohali for chemical analysis.  YS acknowledges DST, India for support through Ramanujan Grant \#SR/S2/RJN-76/2010 and through DST grant \#SB/S2/CMP-001/2013. KM acknowledges UGC-CSIR India for a fellowship.

\end{document}